\documentclass[twocolumn,showpacs,preprintnumbers,amsmath,amssymb]{revtex4}

\usepackage{epsfig}

\begin{document}

\title{Parametrized canonical transformation for the Hubbard-model at
  arbitrary interaction strength} 

\author{ Bal\'azs Het\'enyi and Hans Gerd Evertz} 

\address{ Institut f\"ur Theoretische Physik, TU Graz, 8010 Graz,
  Austria }

\begin{abstract}
  The $t-J$ and Heisenberg models are truncated expansions of a canonically
  transformed Hubbard model coinciding with it at $U\rightarrow \infty$.  We
  show that a modified canonical transformation applied to the Hubbard model
  leads to alternative models of similar form, but whose convergence
  properties with respect to the expansion are more favourable, resulting in
  a good description of the half-filled ground state even at $0<U\leq1$.  We
  investigate the transformed Hamiltonian and observables for metallic and
  insulating variational wave-functions.
\end{abstract}

\maketitle

\label{sec:intro}

The Hubbard model~\cite{Hubbard63,Kanamori63,Gutzwiller63,Gutzwiller65} and
its descendants have contributed greatly to our understanding of strongly
correlated systems~\cite{Imada98,Auerbach98,Fazekas99} and in particular the
metal-insulator transition~\cite{Imada98} (MIT) exhibited by these systems.
Early attempts~\cite{Gutzwiller65,Brinkman70} to explain the MIT were based
on the use of a projected wavefunction due to Gutzwiller (GW). An approximate
variational calculation based on the Gutzwiller approximation
(GA)~\cite{Gutzwiller65} for the GW in the general case predicts a
MIT~\cite{Brinkman70} between a paramagnetic metal and an insulator
(Brinkman-Rice transition).  The order parameter for the Brinkman-Rice
transition is the fraction of doubly occupied sites which goes to zero at the
critical $U=U_c$.  A shortcoming of the GA is that second-order hopping
processes are not included, i.e. double occupations that arise as a result of
second-order hoppings (which give rise to anti-ferromagnetic (AFM) coupling)
are entirely absent.  Thus the number of double occupations is not a valid
order parameter for the actual MIT.  In one dimension, the exact solution for
the Hubbard model~\cite{Lieb68} indicates insulating behavior for all finite
values of the interaction, whereas the exact solution for the
GW~\cite{Metzner87} for the same system is always metallic.

The importance of higher-order hopping processes is made obvious by a
canonical transformation of the Hubbard model that eliminates those
first-order hopping processes which increase(decrease) the number of doubly
occupied sites
($H_t^+$($H_t^-$))~\cite{Kohn64,Harris67,Emery76,Chao78,Gros87,MacDonald88}.
Expansion and truncation of the transformed Hamiltonian leads to the
well-known $t-J$ and spin$-\frac{1}{2}$ anti-ferromagnetic Heisenberg models,
which coincide with the Hubbard model in the strong-coupling limit, in which
it leads to anti-ferromagnetism.~\cite{Gros87,Scalapino06}

The effective Hamiltonians derived from the Hubbard model have other
applications as well.  In the resonating valence bond (RVB)
method~\cite{Anderson73,Fazekas74,Anderson04} the expectation value of the
$t-J$ Hamiltonian is evaluated over a fully Gutzwiller projected
wavefunction~\cite{Gutzwiller63,Gutzwiller65}.  The RVB wavefunction has
recently been applied to the problem of high temperature superconductivity,
and many experimentally observed features of the relevant materials have been
reproduced.~\cite{Anderson04,Paramekanti01,Edegger06} 

In the present study the unitary operator that transforms the Hubbard model
into the $t-J$ or Heisenberg models is parametrized so that the number of
double occupations as a function of the transformation can be minimized.  The
effect of our procedure is similar to that of the original transformation.
The difference is that $H_t^+$ and $H_t^-$ are not {\it cancelled} from the
Hamiltonian as in the standard case, but instead {\it constrained} so that
their expectation values are zero.  In contrast, the $t-J$ and
spin-$\frac{1}{2}$ Heisenberg models will in general give finite expectation
values for $H_t^+$ and $H_t^-$.  In our approach first-order double
occupations are eliminated at the wavefunction level, as opposed to the
operator (Hamiltonian) level.  The optimized transformation can be applied at
any value of the interaction and not only in the strongly interacting limit.
We diagonalize the transformed Hamiltonians for systems of up to $12$ lattice
sites, and it is shown that the optimized expansion converges much faster
than the standard one.  Convergence is also demonstrated for $U\leq1$.

We also investigate the behavior of the optimally transformed double
occupation operator using two different variational wavefunctions the
Gutzwiller~\cite{Gutzwiller65} (GW) and Baeriswyl~\cite{Baeriswyl86} (BW)
wavefunctions and compare them to the exact result.

The Hubbard model Hamiltonian can be written as
\begin{equation}
  H = \overbrace{-t \sum_{\langle i,j \rangle \sigma}
    c_{i\sigma}^{\dagger}c_{j\sigma}}^{H_t} + 
  \overbrace{UD}^{H_U}
\end{equation}
where $D = \sum_{i} n_{i \uparrow}n_{i \downarrow}$ and where the operator
$c_{i\sigma}^{\dagger}$($c_{i\sigma}$) creates(destroys) a particle at site
$i$ with spin $\sigma$, and $n_{i\sigma}$ is the density operator at site $i$
for particles of spin $\sigma$.  In deriving the canonically transformed
Hamiltonian it is helpful to break up the kinetic energy operator into terms
consisting of different types of hoppings~\cite{Fazekas99}:
\begin{equation}
  H_t = H_t^+ + H_t^- + H_t^0,
\end{equation}
where
\begin{eqnarray}
H_t^+&=& -t \sum_{\langle i,j \rangle \sigma} n_{i-\sigma}c_{i\sigma}^{\dagger}c_{j\sigma}(1-n_{j-\sigma})  \\ 
H_t^0&=& -t \sum_{\langle i,j \rangle \sigma} n_{i-\sigma}c_{i\sigma}^{\dagger}c_{j\sigma}n_{j-\sigma} \nonumber  \\
&&  -t \sum_{\langle i,j \rangle \sigma}
(1-n_{i-\sigma})c_{i\sigma}^{\dagger}c_{j\sigma}(1-n_{j-\sigma}) \nonumber \\
H_t^-&=& -t \sum_{\langle i,j \rangle \sigma} (1-n_{i-\sigma})c_{i\sigma}^{\dagger}c_{j\sigma}n_{j-\sigma}. \nonumber
\end{eqnarray}
$H_t^+$($H_t^-$) include only hopping processes which
increase(decrease) the number of double occupations, and $H_t^0$ includes
only those which leave the number of double occupations unchanged.  The
Hermitian operator defined as
\begin{equation}
S = -\frac{i}{U}(H_t^+-H_t^-)
\end{equation}
is useful in defining the transformation
\begin{equation}
  H_{S} = e^{iS} H e^{-iS} = H + i[S,H] + \frac{i^2}{2}[S,[S,H]] + ...
\label{eqn:H_eff}
\end{equation}
The series can be viewed as a power series in $\frac{t}{U}$.  It can be shown
that
\begin{equation}
i[S,H_U] = -(H_t^+ + H_t^-),
\label{eqn:comSHU}
\end{equation}
and thus, up to first order, hoppings that change the number of double
occupations are {\it cancelled} from the transformed Hamiltonian (Eq.
(\ref{eqn:H_eff})).  The $t-J$ and Heisenberg models, which are used as
effective models in the large $U$ limit, can be derived by explicitly
evaluating the terms of Eq. (\ref{eqn:H_eff}) up to second order in $t/U$,
\begin{eqnarray}
\label{eqn:H_eff_app}
H_S \approx &H_t^0 + H_U + J \sum_{\langle i,j \rangle} \left( {\bf S}_i \cdot
  {\bf S}_j - \frac{n_i n_j}{4} \right) \\ & + \mbox{3-site terms}\nonumber
\end{eqnarray}
where $J = 4t^2/U$.
\begin{figure}[htp]
\vspace{1cm}
\psfig{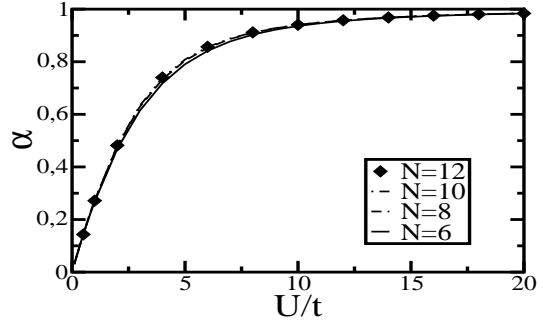}
\vspace{1cm}
\caption{Optimal $\alpha$ as a function of $U$ for systems with different
  sizes.}
\label{fig:alpha}
\end{figure}

We now consider a similar transformed Hamiltonian derived using the modified
operator $e^{i\alpha S}$ which leads to
\begin{eqnarray}
\label{eqn:Hb_eff}
H_{\alpha S} &= & e^{i\alpha S} H e^{-i \alpha S}\\ &=& H + i \alpha [S,H] +
\frac{i^2\alpha^2}{2}[S,[S,H]] + ... , \nonumber
\end{eqnarray}
where $\alpha$ is a parameter to be determined.  If for a particular state
the transformed number of double occupations
\begin{equation}
  \langle \Psi | D_{\alpha S} | \Psi \rangle = \langle \Psi | e^{i\alpha S} D
  e^{-i\alpha S} | \Psi \rangle,
\label{eqn:d}
\end{equation}
is minimized as a function of $\alpha$, then it holds that
\begin{equation}
\langle \Psi | e^{i\alpha S} [S,D] e^{-i\alpha S} | \Psi \rangle = 0,
\label{eqn:Ndeq0}
\end{equation}
which with Eq. (\ref{eqn:comSHU}) is equivalent to
\begin{equation}
\langle \Psi | e^{i\alpha S} (H_t^+ + H_t^-) e^{-i\alpha S} | \Psi \rangle =
0.
\label{eqn:Htpm_eff}
\end{equation}
Thus, double occupations up to first-order can be excluded via a
transformation that sets the expectation value of the sum of the operators
$H_t^+ + H_t^-$ to zero.  The main difference between the Hamiltonians in
Eq. (\ref{eqn:H_eff}) and Eq. (\ref{eqn:Hb_eff}) is that in the latter the
{\it expectation value} of the sum of the operators that change the number of
double occupations is zero, as opposed to being cancelled by another term
equal but opposite in sign at the operator level.

If $\Phi$ is the ground state of the Hubbard Hamiltonian, then
\begin{equation}
  \langle \Phi | H | \Phi \rangle = \langle \Phi_{\alpha
    S} | H_{\alpha S} | \Phi_{\alpha S} \rangle,
\end{equation}
where the transformed wavefunction $|\Phi_{\alpha S} \rangle = e^{i\alpha S}
|\Phi \rangle$ is the ground state of the transformed Hamiltonian $H_{\alpha
  S}$.  While the optimization procedure can be carried out on any state, in
the rest of this work we deal exclusively with the ground state at half
filling.
\begin{table}
\begin{tabular}{|c||c||c|c|c|c|}
\hline
Hamiltonian & $U$ & 2nd order & 4th order & 6th order & Exact \\
\hline
$e^{iS} H e^{-iS}$   &0.5&-395.505&-4613.096&-35947.499&-7.275\\
   &1.0 &-99.211&-270.19&-495.743&-6.601\\
   &2.0 &-24.713&-10.3987&-7.6983&-5.409\\
   &5.0 &-4.557&-2.974&-3.092&-3.088\\
   &10.0&-1.824&-1.661&-1.664&-1.664\\
\hline
$e^{i\alpha S} H e^{-i\alpha S}$   &0.5&-11.084&-6.695&-7.328&-7.275\\
   &1.0&-9.850&-6.159&-6.634&-6.601\\
   &2.0&-7.742&-5.158&-5.421&-5.409\\
   &5.0&-3.819&-3.047&-3.088&-3.088\\
   &10.0&-1.792&-1.662&-1.664&-1.664\\
\hline
\end{tabular}
\caption{
  Comparison of ground state energies calculated for a lattice composed of six
  sites.  The upper(lower) half shows results for the transformed Hamiltonian
  with $\alpha=1$(optimized $\alpha$). The rightmost column shows the exact
  results.  The expansion is in the parameter $\alpha$.
}
\label{tab:energy}
\end{table}

The analog derivation that leads to the $t-J$ model applied to
Eq. (\ref{eqn:Hb_eff}) results in
\begin{eqnarray}
\label{eqn:Hb_eff_app}
H_{\alpha S} \approx & H_t^0 + H_U + J_{\alpha S} \sum_{\langle i,j \rangle}
\left( {\bf S}_i \cdot {\bf S}_j - \frac{n_i n_j}{4} \right) \\ &+ \mbox{3-site terms},\nonumber
\end{eqnarray}
where $J_{\alpha S}$ denotes a modified coupling constant satisfying
\begin{equation}
J_{\alpha S}=(2\alpha - \alpha^2)J. 
\end{equation}
The first-order term in $\alpha$ originates from the transformed
$H_t^+$ and $H_t^-$.

The size of the parameter $\alpha$ determines the convergence of the
expansion (Eq. (\ref{eqn:Hb_eff})).  In Fig. \ref{fig:alpha} the results of
power-method type calculations~\cite{Hetenyi08} are shown for systems of
various sizes at half-filling.  Anti-periodic(periodic) boundary conditions
were applied for system sizes with odd(even) multiples of
two~\cite{Stafford91,Fye91}.  The parameter $\alpha$ which minimizes Eq.
(\ref{eqn:d}) ( and satisfies Eqs.  (\ref{eqn:Ndeq0}) and
(\ref{eqn:Htpm_eff})) and is closest to the origin is calculated as a
function of the interaction parameter $U$.  We find that convergence is
achieved for all $U$ considered.  As expected, $H_S$ is recovered for large
$U$.  The size-dependence of $\alpha$ is negligible.  Interestingly, as $U$
approaches zero $\alpha/U$ converges to $\approx 0.3$, wheras in the standard
case $1/U$ diverges.
\begin{figure}[htp]
\vspace{1cm}
\psfig{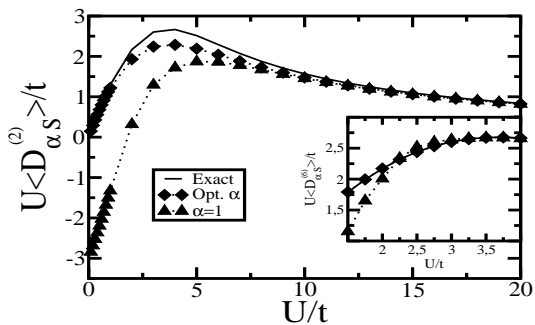}
\vspace{1cm}
\caption{Ground state expectation value of the transformed interaction U
  $\langle D_{\alpha S} \rangle/t$ for the standard expansion and the
  optimized one compared to the exact results for a model with six sites.
  The transformed $D_{\alpha S}$ was expanded to second order in $\alpha$ in
  the main plot, sixth order in the inset.}
\label{fig:H_U}
\end{figure}

In Table \ref{tab:energy} we compare energies calculated using the standard
transformation (Eq. (\ref{eqn:H_eff})), and those resulting from the
transformation with optimized $\alpha$ (Eq. (\ref{eqn:Hb_eff}) and Fig.
\ref{fig:alpha}).  The optimal value of $\alpha$ was obtained from exact
diagonalization.  In these calculations periodic boundary conditions were
used.  Subsequently, $\alpha$ was used in the expansion, Eq.
(\ref{eqn:Hb_eff}).  In order to investigate the convergence, the expansion
of the Hamiltonian was carried out to second, fourth, and sixth orders in
$\alpha$, then diagonalized.  The optimized transformation gives energies
closer to the exact result in all cases, and the convergence is also better
when the expansion of the Hamiltonian is carried out to higher orders.  The
advantage is more pronounced at lower values of $U$, in particular our
transformation is even applicable for $U\leq1$ where the standard expansion
fails due to slow convergence.  The second order results with optimal
$\alpha$ (similar to the $t-J$ model) are in considerably better agreement
with the exact results than the standard ($\alpha=1$) second order ones,
therefore the $t-J$ model is, in this sense, applicable even at $U\leq1$, but
with a modified coupling.
\begin{figure}[htp]
\vspace{1cm}
\psfig{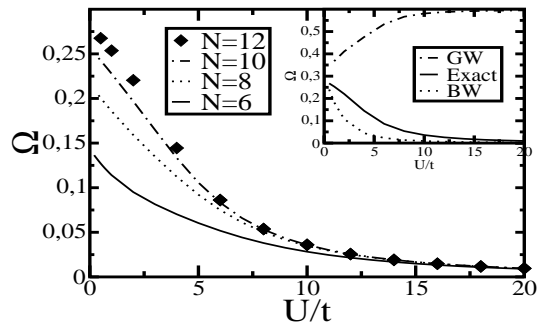}
\vspace{1cm}
\caption{Ratio $\Omega$ (defined in Eq. (\ref{eqn:omega})) calculated exactly
  for different system sizes.  The inset shows a comparison between the exact
  result and two different variational wavefunctions (Baeriswyl (BW) and
  Gutzwiller (GW)) for the system with $12$ lattice sites.  }
\label{fig:doc}
\end{figure}

In Fig. \ref{fig:H_U} the expectation value of the transformed interaction
energy is shown.  The expansion is carried out to second and sixth order
(inset) for $\alpha=1$ and for optimized $\alpha$, i.e. the Haniltonian is
calculated up to a given order, and diagonalized.  The observable is also
transformed and truncated at the given order.  Optimized $\alpha$ gives
quantitative agreement with the exact result even at second order ($t-J$ like
model), whereas the standard version is not in agreement with the exact
results at second order, and even when the expansion is carried out to sixth
order, agreement is only reached when $U$ is large.

The $t-J$ type model derived herein is not as easy to derive as the standard
one.  At a particular $U$ the normal $t-J$ model can easily be derived to any
order.  Our modified model depends on a parameter, $\alpha$, which is a
function of the ground state solution.  For a particular $U$ one can obtain
$\alpha$ by expanding the transformed Hamiltonian (Eq. (\ref{eqn:Hb_eff})),
solving for its ground state, and varying $\alpha$ to satisfy the condition
in Eq. (\ref{eqn:Ndeq0}).  It also appears possible to apply our formalism
using the generalized version of the canonical transformation of
Ref. \cite{MacDonald88}.

We have also investigated our scheme for different variational wavefunctions.
For our studies we have chosen the Baeriswyl and Gutzwiller wavefunctions (BW
and GW respectively).  The properties of these wavefunctions are well-known.
In particular it has been shown by Millis and Coppersmith~\cite{Millis91}
that the Drude weight of the GW is always finite in the thermodynamic limit,
hence the GW is metallic.  This property can be attributed to the lack of
explicit phase dependence of the GW.  The BW has been shown to consist of
rotating dipoles formed of empty and doubly occupied sites, and to be in
general an insulating wavefunction~\cite{Baeriswyl00}.

In Fig.  \ref{fig:doc} we present a comparison of the ratio
\begin{equation}
  \Omega = \frac{\langle \Psi | D_{\alpha S} | \Psi \rangle }{\langle \Psi | D | \Psi \rangle }
\label{eqn:omega}
\end{equation}
for systems with different sizes calculated exactly.  As $U$ increases
$\Omega$ decreases sharply.  The inset in Fig.  \ref{fig:doc} shows a
comparison for the system of size $12$ between the exact result and two
variational wavefunctions BW and GW.  An interesting feature is that in the
large $U$ limit the GW tends to a finite value unlike the exact or the BW
result.  These qualitative tendencies persist away from half-filling (results
not shown).  Hence the GW tending to a finite limit is not due to
metallicity.
\begin{figure}[htp]
\vspace{1cm}
\psfig{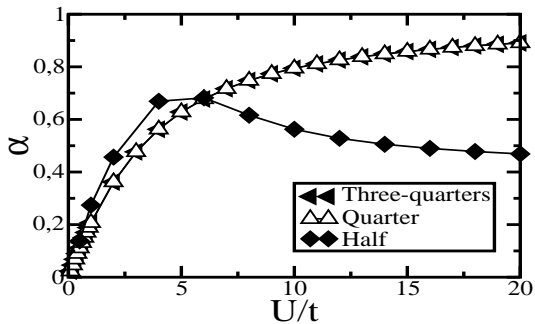}
\vspace{1cm}
\caption{Optimal $\alpha$ as a function of $U$ for the Gutzwiller
  wavefunction with $12$ sites for different fillings.}
\label{fig:alpha_gutz}
\end{figure}

In Fig. \ref{fig:alpha_gutz} we show the optimum $\alpha$ at three different
fillings for the GW.  At half-filling the behaviour is qualitatively
different from the other fillings investigated, and different from the
behaviour found for the exact case (Fig. \ref{fig:alpha}).  At large $U$
$\alpha$ appears to be bounded below for half-filling, where GW is expected
to be in error, since it is a metallic wavefunction.  Away from half-filling
the $\alpha$ obtained from GW is monotonically increasing.  We have also
investigated the BW and found the qualitative tendencies (monotonic increase,
upper bound of $\alpha=1$) to be the same as for the exact calculation.

In conclusion we have shown that the standard canonical transformation which
when applied to the Hubbard model gives the $t-J$ model at large interaction
strength can be optimized to give a $t-J$ like model applicable for the whole
range of the interaction strength.  In particular convergence of the expanded
Hamiltonian is achieved for interaction strength close to zero, where the
standard transformation leads to slow convergence.

Beneficial discussions with E. Arrigoni and W. von der Linden are gratefully
acknowledged.



\end{document}